\documentstyle[12pt,a4wide]{article}
\input epsf

\newcommand{\news}{\setcounter{equation}{0}\ \indent}
\newcommand{\be}{\begin{equation}}
\newcommand{\ee}{\end{equation}}
\newcommand{\bea}{\begin{eqnarray}}
\newcommand{\eea}{\end{eqnarray}}
\newcommand{\bean}{\begin{eqnarray*}}
\newcommand{\eean}{\end{eqnarray*}}
\font\upright=cmu10 scaled\magstep1
\font\sans=cmss12
\newcommand{\ssf}{\sans}
\newcommand{\stroke}{\vrule height8pt width0.4pt depth-0.1pt}
\newcommand{\Z}{\hbox{\upright\rlap{\ssf Z}\kern 2.7pt {\ssf Z}}}

\newcommand{\C}{{\rlap{\rlap{C}\kern 3.8pt\stroke}\phantom{C}}}
\newcommand{\R}{\hbox{\upright\rlap{I}\kern 1.7pt R}}
\newcommand{\CP}{\C{\upright\rlap{I}\kern 1.7pt P}}
\newcommand{\half}{\frac{1}{2}}
\newcommand{\mt}{\rlap{\ssf T}\kern 3.0pt{\ssf T}}
\newcommand{\spc}{spectral curve }

\newcommand{\identity}{{\upright\rlap{1}\kern 2.0pt 1}}

\newcommand{\wpp}{{\wp^\prime}}
\newcommand{\dop}{(\zeta_0\frac{\partial}{\partial\zeta_1})}
\newcommand{\ady}{(\mbox{ad}Y)}
\begin{document}
\pagestyle{plain}

\title{\vskip -70pt
\begin{flushright}
{\normalsize DAMTP 95-20} \\
{\normalsize To appear in Nonlinearity} \\
\end{flushright}
\vskip 20pt
{\bf Octahedral and Dodecahedral Monopoles} \vskip 10pt}

\author{Conor J. Houghton \\[10pt]
and \\[10pt]
Paul M. Sutcliffe\thanks{
Address from September 1995,
 Institute of Mathematics,
University of Kent at Canterbury, Canterbury CT2 7NZ. 
Email P.M.Sutcliffe@ukc.ac.uk
} \\[20pt]
{\sl Department of Applied Mathematics and Theoretical Physics} \\[5pt]
{\sl University of Cambridge} \\[5pt]
{\sl Silver St., Cambridge CB3 9EW, England}\\[5pt]
{\normalsize c.j.houghton@damtp.cam.ac.uk \& p.m.sutcliffe@damtp.cam.ac.uk}
 \\[10pt]}

\date{May 1995}
\maketitle
\begin{abstract}
It is shown that there exists a charge five monopole with octahedral symmetry and a
charge seven monopole with icosahedral symmetry. A numerical
implementation of the ADHMN construction is used to calculate 
the energy density of these monopoles and
surfaces of constant energy density are displayed.
 The charge five and charge seven monopoles look like an
octahedron and a dodecahedron respectively. A scattering geodesic for
each of these monopoles is presented and discussed 
using rational maps.
This is done with the aid of a new formula for the cluster
decomposition of monopoles when the poles of the rational map are
close together.
\end{abstract}
\newpage
\section{Introduction}
\news
BPS monopoles are topological solitons in a three dimensional SU(2)
Yang-Mills-Higgs gauge theory, in the limit of vanishing Higgs potential.
They are solutions to the Bogomolny equation
\be D_A\Phi=\star F_A
\label{bogeq}\ee
where $D_A$ is the covariant
derivative, with $A$ an $su(2)$-valued gauge potential 1-form, $F_A$
its gauge
field 2-form and $\star$ the Hodge dual on \R$^3$.
 The Higgs field, $\Phi$, is an $su(2)$-valued
scalar field and is required to satisfy
\be\|\Phi\|\stackrel{r\rightarrow\infty}{\longrightarrow}1\label{bcond}\ee
where $r= | x | $ and $\|\Phi\|^2=-\half\mbox{tr}\Phi^2$. The boundary
condition (\ref{bcond}) can be considered to be a residual finite energy
condition, derived from the now vanished Higgs potential.

The Higgs field at infinity induces a map between spheres:
\be\Phi:S^2(\infty)\rightarrow S^2(1)\ee
where $S^2(\infty)$ is the two-sphere at spatial infinity and
$S^2(1)$ is the two-sphere of vacuum configurations given by
$\{\Phi\in su(2) : \|\Phi\|=1\}.$
The degree of this map is a non-negative integer $k$ which
(in suitable units) is the total magnetic charge. 
We shall refer to a monopole
with magnetic charge $k$ as a $k$-monopole. 
The total energy of a $k$-monopole is equal to $8\pi k$ and the energy
density may be expressed \cite{Wa} in the convenient form
\be
{\cal E}=\bigtriangleup \|\Phi\|^2
\label{lap}
\ee
where $\bigtriangleup$ denotes the laplacian on \R$^3$.

Monopoles correspond to certain algebraic curves, called spectral
curves,
 in the mini-twistor
space \mt$\cong$T\CP$^1$ \cite{Wa,HA,HB}. This space is isomorphic to the
space of directed lines in $\R^3$. If $\zeta$ is the standard
inhomogeneous coordinate on the base space, it corresponds
 to the direction of a line in $\R^3$.
The fibre coordinate, $\eta$, 
is a complex coordinate in a plane orthogonal to this line.
The spectral curve of a monopole
is the set of lines along which the differential equation 
\be (D_A-i\Phi)v=0\label{de}\ee
has bounded solutions in both directions.
The spectral curve of a $k$-monopole takes the form
\be\eta^k+\eta^{k-1} a_1(\zeta)+\ldots+\eta^r a_{k-r}(\zeta)
+\ldots+\eta a_{k-1}(\zeta)+a_k(\zeta)=0\ee
where, for $1\leq r\leq k$, $a_r(\zeta)$ is a polynomial in $\zeta$ of
maximum degree $2r$.
However, general curves of this form will only correspond 
to $k$-monopoles if they
satisfy the reality condition
\be a_r(\zeta)=(-1)^r\zeta^{2r}\overline{a_r(-\frac{1}{\overline{\zeta}})}\label{reality}\ee
and some difficult non-singularity conditions \cite{HA}. In  \cite{HMM} the
concept of a strongly centred monopole is introduced. A strongly
centred monopole is centred on the origin and its rational map
has total phase one.
If a
monopole is strongly centred its spectral curve satisfies
\be a_1(\zeta)=0\label{sc}.\ee

Even though the Bogomolny equation is integrable, it is not easily
solved. Explicit solutions are only known in the cases of
1-monopole \cite{PS}, 2-monopoles \cite{Wa,Wb} and axisymmetric monopoles of
higher charges \cite{P}. Recently, progress has been made in
understanding multi-monopoles. 
Hitchin, Manton and Murray \cite{HMM} have demonstrated the
existence of monopoles corresponding to the spectral curves
\begin{eqnarray}\eta^3+i\frac{\Gamma(1/6)^3\Gamma(1/3)^3}{48\sqrt{3}\pi^{3/2}}\zeta(\zeta^4-1)&=&0\label{tet}\\
\eta^4+\frac{3\Gamma(1/4)^8}{64\pi^2}
(\zeta^8+14\zeta^4+1)&=&0\label{oct}.\end{eqnarray}
The first spectral curve (\ref{tet}) has tetrahedral symmetry, the
second (\ref{oct}) has octahedral symmetry. In \cite{HS} we computed
numerically and displayed surfaces of constant energy density for 
these monopoles. We noted that the charge four monopole looks like
a cube, rather than an octahedron. We therefore refer to this 
4-monopole as a cubic monopole.

Hitchin, Manton and Murray \cite{HMM} also prove that although 
\be\zeta_1^{11} \zeta_0+11\zeta_1^6\zeta_0^6-\zeta_1\zeta_0^{11}\ee
is an icosahedrally invariant homogeneous polynomial of degree 12,
the invariant algebraic curve 
\be\eta^6+a\zeta(\zeta^{10}+11\zeta^5-1)=0\ee
does not correspond to a monopole for any value of $a$. 
However, based upon considerations of the symmetries of rational
maps for infinite curvature hyperbolic monopoles, Atiyah has
suggested, \cite{A} \footnote{We thank
 Nick Manton for drawing 
this to our attention}
 that there may be an icosahedrally
invariant 7-monopole.
In this paper, we prove that this suggestion is correct by demonstrating
that the algebraic curve
\be\eta^7+
\frac{\Gamma(1/6)^6\Gamma(1/3)^6}{64\pi^3}
\zeta(\zeta^{10}+11\zeta^5-1)\eta=0
\label{dodsc}\ee 
is the spectral curve of a monopole. Using our numerical scheme 
introduced in \cite{HS}, we then compute its energy density. 
On examining surfaces of constant
energy density, we find that the charge seven monopole looks like a
dodecahedron. 

In each of the cases examined so far, the minimum charge
monopole with the symmetry of a regular solid has charge $k=\half(F+2)$,
where $F$ is the smallest number of faces of a regular solid with
that symmetry. This leads us to conjecture that the minimum charge
monopole resembling a regular solid with $F$ faces has charge
$k=\half(F+2)$. 
For the dodecahedron $F=12$, which gives $k=7$. In fact, this
conjecture was one of the motivations for our consideration of charge
seven when searching for an icosahedrally symmetric monopole.
In this paper, we demonstrate that our conjecture is also correct for the
octahedron by proving that the octahedrally symmetric algebraic curve
\be \eta^5+\frac{3\Gamma(\frac{1}{4})^8}{16\pi^2}(\zeta^8+14\zeta^4+1)\eta=0\ee
is the spectral curve of a 5-monopole.
We display its energy density and confirm that
 it looks like an octahedron. It remains to be
verified that an icosahedrally symmetric monopole of charge eleven
exists and resembles an icosahedron.

It is interesting that numerical evidence suggests that 
similar results hold in the case of static minimum energy
multi-skyrmion solutions. In \cite{BTC} Braaten, Townsend and Carson
use a discretization of the Skyrme model on a cubic lattice to
calculate such solutions for baryon numbers $B=3,4,5$ and $6$. They
find that surfaces of constant baryon number density resemble solids with
$2B-2$ faces. 
Furthermore, the fields describing solutions with $B=3$ and $B=4$ are
seen to possess tetrahedral and octahedral symmetry. However, 
they conclude that the solution for
$B=5$ seems only to have $D_{2d}$ symmetry. This contrasts with the
existence of a charge five monopole with octahedral symmetry.

Approximations to the $B=3$ and $B=4$ skyrmions have been calculated
by computing the holonomies of Yang-Mills instantons \cite{LM}.
These instanton generated Skyrme fields also have tetrahedral
and octahedral symmetry respectively.
Given the numerical evidence for an apparent difference between
charge five monopoles and skyrmions, it would
be instructive to construct instanton-generated Skyrme fields
with baryon number five. It may be that an octahedrally symmetric
5-skyrmion 
simply does not exist. However, the instanton construction could shed
some light on other possibilities; for example, that such a skyrmion
exists but it does not have minimum energy. A second possibility is
that the numerical scheme used in \cite{BTC} is responsible for no
such skyrmion being found. For particular orientations, an octahedron
will not fit inside a cubic lattice; in the sense of all the vertices
of the octahedron sitting on lattice sites. The discretization could
then result in the octahedron being squashed into a shape similar to that found in \cite{BTC}. Of course, at the moment, all these
possibilities are pure speculation. What is clear from our results is
that the $B=7$ skyrmion should now be investigated, as there is some
interest in the possibility that this is icosahedrally symmetric.

In Section 2, we outline the ADHMN construction as applied to  symmetric
monopoles. In Sections 3 and 4, we present our results
on dodecahedral and octahedral monopoles.
Finally, in Section 5, we discuss rational maps
and geodesic monopole scattering related to these symmetric monopoles.
This is done with the aid of a new formula for the cluster
decomposition of monopoles when the poles of the rational map are
close together.

\section{The Nahm Equations}
\news
The main difficulty in proving that an algebraic curve is the spectral
curve of a
monopole lies in demonstrating satisfaction of the non-singularity
conditions. However, there is a reciprocal formulation of the
Bogomolny equation in which non-singularity is manifest. This
formulation is the
Atiyah-Drinfeld-Hitchin-Manin-Nahm
 (ADHMN) construction \cite{N,HB}. This is an equivalence
 between $k$-monopoles and Nahm data
$(T_1,T_2,T_3)$, which are three $k\times k$ matrices depending
on a real parameter $s\in[0,2]$ and satisfying:\\
\\
\newcounter{con}
\setcounter{con}{1}
(\roman{con})  Nahm's equation
\be
\frac{dT_i}{ds}=\half\epsilon_{ijk}[T_j,T_k] \label{Neqn}
\ee\\
\\
\addtocounter{con}{1}
(\roman{con}) $T_i(s)$ is regular for $s\in(0,2)$ and has simple
poles at $s=0$ and $s=2$,\\
\\
\addtocounter{con}{1}
(\roman{con}) the matrix residues of $(T_1,T_2,T_3)$ at each
pole form the irreducible $k$-dimensional representation of SU(2),\\
\\
\addtocounter{con}{1}
(\roman{con}) $T_i(s)=-T_i^\dagger(s)$,\\
\\
\addtocounter{con}{1}
(\roman{con}) $T_i(s)=T_i^t(2-s)$.\\

It should be noted that in this paper we shall not search
for a basis in which property (\roman{con}) is explicit, but
rely on a general argument that such a basis
exists (see \cite{HMM}).

Explicitly, the \spc may be read off from the Nahm data as the
equation
\be
\mbox{det}(\eta+(T_1+iT_2)-2iT_3\zeta+(T_1-iT_2)\zeta^2)=0.
\label{curve}
\ee
It is obvious from (\ref{curve}) that the strong centering condition
(\ref{sc}) is equivalent to\\
\\
\addtocounter{con}{1}
(\roman{con}) $\mbox{tr}T_i(s)=0$.\\

\setcounter{con}{1}

To extract the monopole fields $(\Phi,A)$ from the Nahm data 
requires the computation of a basis for the kernel of a linear
differential operator constructed out of the Nahm data, followed
by some integrations. We have developed a numerical algorithm
which can perform all these required tasks, the details are included
in \cite{HS}. The algorithm takes as
input the Nahm data and outputs the energy density of the 
corresponding monopole. It will be applied to the Nahm data
which we construct in this paper.

As in \cite{HS} we use the discrete symmetry group $G$ of 
the conjectured monopole
to reduce the number of Nahm equations. Since the Nahm matrices are
traceless, they transform under the rotation group as
\begin{eqnarray}\underline{3}\otimes 
sl(\underline{k}) &\cong&\underline{3}\otimes
 (\underline{2k-1}\oplus\underline{2k-3}\oplus
 \ldots \oplus \underline{3})\nonumber\\
&\cong&( \underline{2k+1}_u\oplus \underline{2k-1}_m\oplus
\underline{2k-3}_l)\oplus \ldots\nonumber\\ 
&\ldots& \oplus ( \underline{2r+1}_u\oplus \underline{2r-1}_m\oplus
\underline{2r-3}_l)\oplus\ldots\oplus (
\underline{5}_u\oplus \underline{3}_m\oplus\underline{1}_l)\label{decomp}
\end{eqnarray}
where $\underline{r}$ denotes the unique irreducible r dimensional
representation of $su(2)$ and the subscripts $u,m$ and $l$ 
(which stand for upper, middle and lower) are a
convenient notation 
allowing us to distinguish between $2r+1$ dimensional
representations occuring as 
\begin{eqnarray*}\underline{3}\otimes\underline{2r-1}&\cong&\underline{2r+1}_u\oplus \underline{2r-1}_m\oplus
\underline{2r-3}_l,\\ \underline{3}\otimes\underline{2r+1}&\cong&\underline{2r+3}_u\oplus \underline{2r+1}_m\oplus
\underline{2r-1}_l\end{eqnarray*}
and 
$$\underline{3}\otimes\underline{2r+3}\,\,\,\cong\,\,\,\underline{2r+5}_u\oplus \underline{2r+3}_m\oplus
\underline{2r+1}_l.$$
   
We can then use invariant homogeneous polynomials over \CP$^1$ to
construct $G$-invariant Nahm triplets. The vector space of degree $2r$ homogeneous polynomials
$a_{2r}\zeta_1^{2r}+a_{2r-1}\zeta_1^{2r-1}\zeta_0+\ldots+a_0\zeta_0^{2r}$ is the carrier space for $\underline{2r+1}$
under the identification
\be X=\zeta_1 \frac{\partial}{\partial \zeta_0};\qquad Y=\zeta_0
\frac{\partial}{\partial \zeta_1};\qquad H=-\zeta_0
\frac{\partial}{\partial \zeta_0}+\zeta_1 \frac{\partial}{\partial
\zeta_1}.\ee
where $X,Y$ and $H$ are the basis of $su(2)$ satisfying
\be [X,Y]=H,\qquad[H,X]=2X,\qquad[H,Y]=-2Y.\label{HXY}\ee
As explained in \cite{HMM,HS} if $p(\zeta_0,\zeta_1)$ is a $G$-invariant
homogeneous polynomial we can construct a $G$-invariant
$\underline{2r+1}_u$ charge $k$ Nahm triplet by the following scheme. 

\newcounter{sch}
\setcounter{sch}{1}
(\roman{sch}) The inclusion 
\be\underline{2r+1}\hookrightarrow\underline{3}\otimes
  \underline{2r-1}\cong\underline{2r+1}_u\oplus
\underline{2r-1}_m\oplus\underline{2r-3}_l\ee
is given on polynomials by
\be p(\zeta_0,\zeta_1)
\mapsto\xi_1^2\otimes p_{11}(\zeta_0,\zeta_1)+2\xi_0 \xi_1 \otimes p_{10}
(\zeta_0,\zeta_1)+\xi_0^2\otimes 
p_{00}(\zeta_0,\zeta_1)\ee
where we have used the notation
\be p_{ab}(\zeta_0,\zeta_1)=\frac{\partial^2 p}{\partial\zeta_a\partial\zeta_b}(\zeta_0,\zeta_1).\ee

\addtocounter{sch}{1}
(\roman{sch}) The polynomial expression $\xi_1^2\otimes
p_{11}(\zeta_0,\zeta_1)+2\xi_0\xi_1\otimes p_{10}(\zeta_0,\zeta_1)+\xi_0^2\otimes 
p_{00}(\zeta_0,\zeta_1)$ is rewritten in the form
\be\xi_1^2\otimes
q_{11}(\zeta_0\frac{\partial}{\partial\zeta_1})\zeta_1^{2r}
+(\xi_o\frac{\partial}{\partial
  \xi_1})\xi_1^2\otimes
q_{10}(\zeta_0\frac{\partial}{\partial\zeta_1})\zeta_1^{2r}
+\half (\xi_o\frac{\partial}{\partial
  \xi_1})^2\xi_1^2\otimes 
q_{00}(\zeta_0\frac{\partial}{\partial\zeta_1})\zeta_1^{2r}.\ee

\addtocounter{sch}{1}
(\roman{sch}) This then defines a triplet of $k\times k$ matrices. Given a
$k\times k$ representation of $X,Y$ and $H$ above, the invariant Nahm
triplet is given by:
\be (S_1^\prime,S_2^\prime,S_3^\prime)=
(q_{11}(\mbox{ad}Y)X^{r},q_{10}(\mbox{ad}Y)X^{r},q_{00}(\mbox{ad}Y)X^{r}),
\ee
where ad$Y$ denotes the adjoint action of $Y$ and is given on a
general matrix $M$ by ad$YM=[M,Y]$.

\addtocounter{sch}{1}
(\roman{sch}) The Nahm isospace basis is transformed. This
transformation is given by 
\be(S_1,S_2,S_3)=(\half S_1^\prime+S_3^\prime,-\frac{i}{2}
S_1^\prime+iS_3^\prime,-iS_2^\prime).\label{basis}\ee
Relative to this basis the $SO(3)$-invariant Nahm triplet
corresponding to the $\underline{1}_l$ representation in (\ref{decomp})
is given by $(\rho_1,\rho_2,\rho_3)$ where 
\be
\rho_1=X-Y; \qquad \rho_2=i(X+Y); \qquad \rho_3=iH.
\ee

It is also necessary to construct invariant Nahm triplets lying in
the $\underline{2r+1}_m$ representations. To do this, we first
construct the corresponding $\underline{2r+1}_u$ triplet. We then
write this triplet in the canonical form
\begin{eqnarray} [c_0+c_1(\mbox{ad}Y\otimes 1+1\otimes \mbox{ad} Y
  )&+&
\ldots
+c_i(\mbox{ad}Y\otimes 1+1\otimes \mbox{ad}
Y)^i \\ &+&\ldots+c_{2r}(\mbox{ad}Y\otimes 1+1\otimes \mbox{ad} Y
)^{2r}]\,X\otimes X^r\nonumber\label{canon}\end{eqnarray} 
and map this isomorphically into $\underline{2r+1}_m$ by mapping
the highest weight vector $X\otimes X^r$ to the highest weight vector
\be X\otimes \mbox{ad}YX^{r+1}-\frac{1}{r+1}\mbox{ad}YX\otimes X^{r+1}.\ee

\section{Dodecahedral Seven Monopole}
\news
The minimum degree icosahedrally invariant homogeneous polynomial is
\cite{K}
\be
\zeta_1^{11}
\zeta_0+11\zeta_1^6\zeta_0^6-\zeta_1\zeta_0^{11}.\ee
 Polarizing this gives
\be\xi_1^2\otimes(110\zeta_1^9
\zeta_0+330\zeta_1^4\zeta_0^6)+2\xi_1\xi_0\otimes(11\zeta_1^{10}
+396\zeta_1^5\zeta_0^5-11\zeta_0^{11})+
\xi_0^2\otimes(330\zeta_1^6\zeta_0^4-110\zeta_1\zeta_0^9).\ee
This is proportional to
\begin{eqnarray}\xi_1^2\otimes(\zeta_0\frac{\partial}
{\partial\zeta_1}&+&\frac{1}{5040}(\zeta_0\frac{\partial}
{\partial\zeta_1})^6)\zeta_1^{10}+2\xi_1\xi_0\otimes
(1+\frac{1}{840}(\zeta_0\frac{\partial}{\partial\zeta_1})^5
-\frac{1}{10!}(\zeta_0\frac{\partial}{\partial\zeta_1})^{10})
\zeta_1^{10}\nonumber\\&+&\xi_0^2\otimes(\frac{1}{168}
(\zeta_0\frac{\partial}{\partial\zeta_1})^4-\frac{1}{9!}
(\zeta_0\frac{\partial}{\partial\zeta_1})^9)\zeta_1^{10}\end{eqnarray}
which gives matrices
\begin{eqnarray}
  X\otimes(\mbox{ad}Y&+&\frac{1}{5040}(\mbox{ad}Y)^6)X^{5}
+\mbox{ad}YX\otimes(1+\frac{1}{840}(\mbox{ad}Y)^5-\frac{1}{10!}
(\mbox{ad}Y)^{10})X^5\nonumber\\&+&
\frac{1}{2}(\mbox{ad}Y)^2X\otimes
(\frac{1}{168}(\mbox{ad}Y)^4-
\frac{1}{9!}(\mbox{ad}Y)^9)X^5.\label{icosmat}\end{eqnarray}
We choose the basis given by 
\be H=  \left[ 
{\begin{array}{rrrrrrr}
6 & 0 & 0 & 0 & 0 & 0 & 0 \\
0 & 4 & 0 & 0 & 0 & 0 & 0 \\
0 & 0 & 2 & 0 & 0 & 0 & 0 \\
0 & 0 & 0 & 0 & 0 & 0 & 0 \\
0 & 0 & 0 & 0 & -2 & 0 & 0 \\
0 & 0 & 0 & 0 & 0 & -4 & 0 \\
0 & 0 & 0 & 0 & 0 & 0 & -6
\end{array}}
 \right],\ee
$$ Y=\left[ 
{\begin{array}{ccccccr}
0 & 0 & 0 & 0 & 0 & 0 & 0 \\
\sqrt {6} & 0 & 0 & 0 & 0 & 0 & 0 \\
0 & \sqrt {10} & 0 & 0 & 0 & 0 & 0 \\
0 & 0 & \sqrt {12} & 0 & 0 & 0 & 0 \\
0 & 0 & 0 & \sqrt {12} & 0 & 0 & 0 \\
0 & 0 & 0 & 0 & \sqrt {10} & 0 & 0 \\
0 & 0 & 0 & 0 & 0 & \sqrt {6} & 0
\end{array}}
 \right], \;\;
X=  \left[ 
{\begin{array}{rcccccc}
0 & \sqrt {6} & 0 & 0 & 0 & 0 & 0 \\
0 & 0 & \sqrt {10} & 0 & 0 & 0 & 0 \\
0 & 0 & 0 & \sqrt {12} & 0 & 0 & 0 \\
0 & 0 & 0 & 0 & \sqrt {12} & 0 & 0 \\
0 & 0 & 0 & 0 & 0 & \sqrt {10} & 0 \\
0 & 0 & 0 & 0 & 0 & 0 & \sqrt {6} \\
0 & 0 & 0 & 0 & 0 & 0 & 0
\end{array}}
 \right].$$
Using {\scriptsize MAPLE} the invariant Nahm triplet is calculated,
relative to the basis (\ref{basis}), to give the $\underline{13}_u$
invariant
\begin{eqnarray*}
Z_1 & = & \:\; \left[ 
\begin{array}{ccccccc}
0          &5\sqrt {6} & 0              & 0            &
7\sqrt{6}\sqrt {10} & 0 & 0  \\
-5\sqrt{6}& 0          &- 9\sqrt {10}  & 0            & 0                    & 0 & 0 \\
0          &9\sqrt {10}& 0              &5\sqrt {12}  & 0    & 0 &  -7\sqrt {6}\sqrt {10} \\
0 & 0 & -5\sqrt{12}& 0& 5\sqrt {12}& 
0& 0 \\
 - 7\sqrt{6}\sqrt {10}& 0& 0&  - 5\sqrt {12}& 
0&  - 9\sqrt {10}&0 \\
0& 0& 0& 0&9\sqrt {10}& 0& 5
\sqrt {6} \\
0&0&7\sqrt {6}\sqrt {10}& 0&0&  - 
5\sqrt {6}&0
\end{array}
 \right]\\
Z_2 & = & i  \left[ 
\begin{array}{ccccccc}
0&5\sqrt {6}&0&0&-7\sqrt {6}\sqrt {10}&0&0\\
5\sqrt{6}&0&-9\sqrt{10}&0
&0&0&0 \\
0&-9\sqrt{10}&0&5\sqrt{12}
&0&0&7\sqrt {6}\sqrt {10} \\
0&0&5\sqrt {12}&0&5\sqrt {
12}&0&0\\
 -7\sqrt {6}\sqrt {10}& 0& 0& 5\sqrt {
12}& 0&  - 9\sqrt {10}& 0 \\
0&0&0&0&-9\sqrt {10}&0&
5\sqrt {6} \\
0&0&7\sqrt{6} \sqrt {10}&0&0&5\sqrt {6}&0
\end{array}
 \right]\\ 
Z_3 & = & i \left[ 
\begin{array}{ccccccc}
 - 12&0&0&0&-14\sqrt {
6}&0&0 \\
0&48&0&0&0&0&- 14\sqrt {6} \\
0&0&-60&0&0&0&0 \\
0&0&0&0&0&0&0 \\
0&0&0&0&60&0&0 \\
 - 14\sqrt {6}&0&0&0&0&-48& 0
 \\
0&- 14\sqrt {6}&0&0&0&0&12
\end{array}
 \right]
\end{eqnarray*}
To calculate the $\underline{13}_m$ invariant we put (\ref{icosmat})
in the form (\ref{canon}). It is proportional to
\be
[11!(\mbox{ad}Y\otimes1+1\otimes\mbox{ad}Y)
+7920(\mbox{ad}Y\otimes1+1\otimes\mbox{ad}Y)^6-
(\mbox{ad}Y\otimes1+1\otimes\mbox{ad}Y)^{11}]
  X\otimes X^5.
\ee
Then using the isomorphism mentioned earlier we obtain matrices
\begin{eqnarray*}Y_1&=&\:\;\left[ 
\begin{array}{ccccccc}
0&\sqrt {6}&0&0&-\sqrt {6}\sqrt{10}&0&12 \\
\sqrt {6}&0&- 3\sqrt {10}&0&0&12&0 \\
0&-3\sqrt {10}&0&5\sqrt {12}&0&0&-\sqrt {6}\sqrt{10}\\
0&0&5\sqrt {12}&0&-5\sqrt {12}&0&0 \\
-\sqrt {6}\sqrt{10}&0&0&- 5\sqrt {12}&0&3\sqrt {10}&0 \\
0& 12& 0&0&3\sqrt {10}& 0&  - 
\sqrt {6} \\
12&0&-\sqrt {6}\sqrt{10} &0&0&- \sqrt {6}
&0
\end{array}
 \right]\\
Y_2&=&i\left[ 
\begin{array}{ccccccc}
0&\sqrt {6}&0&0&\sqrt {6}\sqrt{10}&0&12 \\
-\sqrt {6}&0&- 3\sqrt {10}&0&0&-12&0 \\
0&3\sqrt {10}&0&5\sqrt {12}&0&0&\sqrt {6}\sqrt{10}\\
0&0&-5\sqrt {12}&0&-5\sqrt {12}&0&0 \\
-\sqrt {6}\sqrt{10}&0&0&5\sqrt {12}&0&3\sqrt {10}&0 \\
0&  12& 0&0&-3\sqrt {10}& 0&  - 
\sqrt {6} \\
-12&0&-\sqrt {6}\sqrt{10} &0&0& \sqrt {6}
&0
\end{array}
 \right]\\
Y_3&=&i\left[ 
\begin{array}{ccccccc}
0&0&0&0&0&- 10\sqrt {6}&0 \\
0&0&0&0&0&0&10\sqrt {6} \\
0&0&0&0&0&0&0 \\
0&0&0&0&0&0&0 \\
0&0&0&0&0&0&0 \\
10\sqrt {6}&0&0&0&0&0&0 \\
0&- 10\sqrt {6}&0&0&0&0&0
\end{array}
 \right].\end{eqnarray*}

In order to derive the reduced Nahm equations we examine, the
commutation relations. The required relations involving
$\rho$ matrices and $Z$ matrices are
\begin{eqnarray}{[}\rho_1,\rho_2]&=&2\rho_3       \nonumber\\
         {[}Z_1,Z_2]&=&-750\rho_3+90Z_3     \label{cri}\\
    {[}Z_1,\rho_2]+{[}\rho_1,Z_2]&=&-10Z_3. \nonumber\end{eqnarray}
Because of the closed form of these relations, it is possible to
derive a consistent set of Nahm equations from the icosahedrally
invariant Nahm data
\be T_i(s)=x(s)\rho_i+z(s)Z_i,\qquad\qquad i \in \{1,2,3\}.\label{ndi}\ee
That is, we can consistently ignore the invariant Nahm triplet
$(Y_1,Y_2,Y_3)$. In fact, if we add $y(s)Y_i$
to (\ref{ndi}), we cannot simultaneously satisfy
$T_i(s)=-T_i^\dagger(s)$ and the reality condition (\ref{reality}) for
non-trivial $y(s)$. Combining (\ref{cri}) and (\ref{ndi})
gives the reduced Nahm equations
\begin{eqnarray}\frac{d x}{ds}&=&2x^2-750z^2\label{n1i}\\
  \frac{dz}{d s}&=&-10xz+90z^2\nonumber\end{eqnarray}
with corresponding spectral curve
\be\eta[\eta^6+a\zeta(\zeta^{10}+11\zeta^5+1)]=0\ee
where
\be a=552960(14xz-175z^2)(x+5z)^4\ee
is a constant.

To solve equations (\ref{n1i}), let $u=x+5z$ and $v=x-30z$ so that
\begin{eqnarray}\frac{du}{ds}&=&2uv\nonumber\\
  \frac{dv}{ds}&=&6u^2-4v^2\nonumber\\ 
a&=&110592(u^6-v^2u^4)\equiv 110592\kappa^6.
\label{eqnu}\end{eqnarray}
Using the constant to eliminate $v$, the equation for $u$ becomes
\be
\frac{du}{ds}=-2u^2\sqrt{1-\kappa^6u^{-6}}.
\ee
If we let $u=-\kappa\sqrt{\wp(t)}$, where $t=2\kappa s$, then
$\wp(t)$ is the Weierstrass function satisfying 
\be{\wpp}^2=4(\wp^3-1)\ee
where, in the above and what follows, primed functions are
differentiated 
with respect to their arguments. Thus the Nahm equations are solved by
\begin{eqnarray}x(s)&=&\frac{2\kappa}{7}\left[-3\sqrt{\wp(2\kappa
      s)}+\frac{{\wpp}(2\kappa s)}{4\wp(2\kappa s)}\right]\\
                z(s)&=&-\frac{\kappa}{35}\left[\sqrt{\wp(2\kappa
                  s)}+\frac{{\wpp}(2\kappa s)}{2\wp(2\kappa s)}\right].\end{eqnarray}

These functions are analytic in $s\in(0,2)$ and have simple poles at
$s=0,2$ provided $\kappa=\omega$, where $2\omega$ is the
real period of $\wp(t)$. Since $\omega$ is explicitly known for this
Weierstrass function, we have
\be\kappa=\frac{\Gamma(1/6)\Gamma(1/3)}{8\sqrt{3\pi\,}}
\ee
and so
\be a=110592\kappa^6=\frac{\Gamma(1/6)^6\Gamma(1/3)^6}{64\pi^3}.\ee

Near $s=0$
\be\wp(2\kappa s)\sim \left (\frac{1}
{2\kappa s} \right ) ^2\ee
and so the residues of $x$ and $z$ are $-1/2$ and $0$
respectively. At $s=2$ they are, respectively, $-5/14$ and $-1/35$. 
At both poles the eigenvalues of the matrix residue of 
$iT_3$ may be calculated and are $\{\pm 3,\pm 2,\pm1,0\}$.
This demonstrates that the  matrix residues
define the irreducible 7-dimensional representation 
at each end of the interval.
Hence, we have proved the existence of a 7-monopole 
with icosahedral symmetry given by the spectral curve
(\ref{dodsc}).
 
The energy density of this monopole is computed using our numerical
implementation of the ADHMN construction. Fig. 1 shows a surface of
constant energy density. This surface
could resemble either an icosahedron or a dodecahedron, 
but, as we remarked earlier, it resembles
the latter. The energy density takes its maximum value
on the 20 vertices of the dodecahedron.

\begin{figure}[ht]
\begin{center}
\leavevmode
\epsfxsize=10cm \epsffile{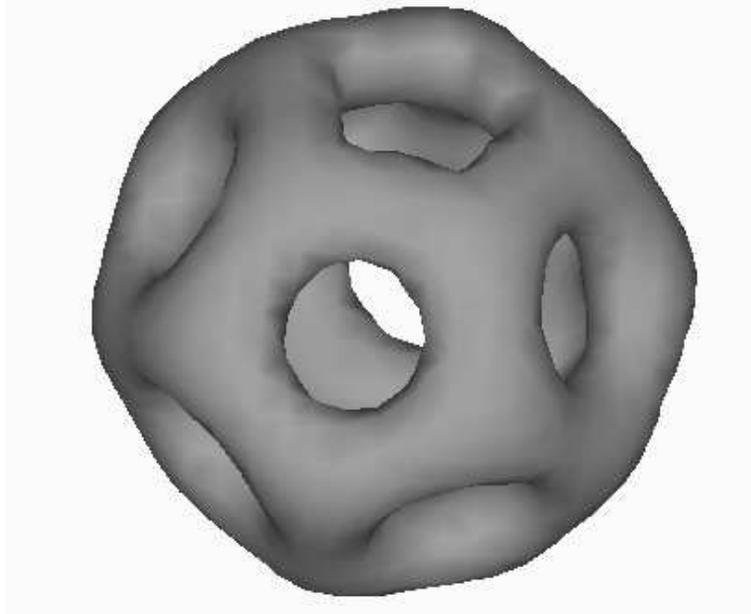}
\caption{Dodecahedral 7-monopole; surface of constant energy density
${\cal E}=0.05$.}
\end{center}
\end{figure}
 
\section{Octahedral Five Monopole}
\news
The lowest degree octahedrally invariant homogeneous polynomial is 
\cite{K}
\be\zeta_1^8+14\zeta_1^4\zeta_0^4+\zeta_0^8.
\label{octhp}\ee 
Polarizing this gives
\be\xi_1^2\otimes(56\zeta_1^6+168\zeta_1^2\zeta_0^4)
+2\xi_1\xi_0\otimes(224\zeta_1^3\zeta_0^3)+
\xi_0^2\otimes(56\zeta_0^6+168\zeta_1^4\zeta_0^2)\ee
which we write in the form
\be\xi_1^2\otimes(56+\frac{7}{15}\dop^4)\zeta_1^6
+2\xi_1\xi_0\otimes\frac{28}{15}\dop^3\zeta_1^6
+\xi_0^2\otimes(\frac{7}{90}\dop^6+\frac{28}{5}\dop^2)\zeta_1^6
\ee
giving matrices
\be
X\otimes(56+\frac{7}{15}\ady^4)X^3
+\ady X\otimes\frac{28}{15}\ady^3X^3
+\frac{1}{2}\ady^2X\otimes(\frac{7}{90}\ady^6+\frac{28}{5}\ady^2)X^3.
\label{nineu}
\ee
If we represent the $su(2)$ basis (\ref{HXY}) by
$$ H =  \left[ 
{\begin{array}{rrrrr}
-4 & 0 & 0 & 0 & 0 \\
0 & -2 & 0 & 0 & 0 \\
0 & 0 & 0 & 0 & 0 \\
0 & 0 & 0 & 2 & 0 \\
0 & 0 & 0 & 0 & 4
\end{array}}
 \right],
X= -i \left[ 
{\begin{array}{ccccr}
0 & 0 & 0 & 0 & 0 \\
 2 & 0 & 0 & 0 & 0 \\
0 &  \sqrt {6} & 0 & 0 & 0 \\
0 & 0 &  \sqrt {6} & 0 & 0 \\
0 & 0 & 0 &  2 & 0
\end{array}}
 \right],
 Y = i \left[ 
{\begin{array}{rcccc}
0 & 2 & 0 & 0 & 0 \\
0 & 0 & \sqrt {6} & 0 & 0 \\
0 & 0 & 0 & \sqrt {6} & 0 \\
0 & 0 & 0 & 0 & 2 \\
0 & 0 & 0 & 0 & 0
\end{array}}
 \right]. $$
this gives the invariant Nahm triplet 
in $\underline9_u$
$$Y_1=i\left[ 
{\begin{array}{ccccc}
0 &  - 6 & 0 & 10 & 0 \\
 - 6 & 0 & 2\,\sqrt {6} & 0 & 10 \\
0 & 2\sqrt {6} & 0 & 2\sqrt {6} & 0 \\
10 & 0 & 2\sqrt {6} & 0 &  - 6 \\
0 & 10 & 0 &  - 6 & 0
\end{array}}
 \right],\;\; 
Y_2=  \left[ 
{\begin{array}{ccccc}
0 & -6 & 0 & -10 & 0 \\
6 & 0 & 2\sqrt {6} & 0 & -10 \\
0 &  -2 \sqrt {6} & 0 & 2\sqrt {6} & 0 \\
10 & 0 &  - 2\sqrt {6} & 0 & -6\\
0 & 10 & 0 & 6 & 0
\end{array}}
 \right], $$
$$Y_3=i\left[ 
{\begin{array}{ccrcc}
8& 0 & 0 & 0 & 0 \\
0 &  - 16& 0 & 0 & 0 \\
0 & 0 & 0 & 0 & 0 \\
0 & 0 & 0 & 16& 0 \\
0 & 0 & 0 & 0 &  - 8
\end{array}}
 \right].$$
The $\underline9_u$ invariant (\ref{nineu}) is written in the form
(\ref{canon}) as
\be
\left[56+\frac{7}{15}(\mbox{ad}Y\otimes 1+1\otimes\mbox{ad}Y)^4
+\frac{1}{720}(\mbox{ad}Y\otimes 1+1\otimes\mbox{ad}Y)^8\right]
X\otimes X^3
\ee
which when mapped using the isomorphism produces
the invariant Nahm triplet in $\underline9_m$ 
$$Z_1= i \left[ 
{\begin{array}{ccccc}
0 &  -1  & 0 &  - 1 & 0 \\
1 & 0&\sqrt {6} & 0 & 1 \\
0 &  - \sqrt {6} & 0 &  - \sqrt {6} & 0 \\
1 & 0 & \sqrt {6} & 0 & 1 \\
0 &  - 1&0 &  - 1 & 0
\end{array}}
 \right],\;\; 
Z_2 =  \left[ 
{\begin{array}{rcccr}
0 & -1 & 0 & 1 & 0 \\
-1 & 0 & \sqrt {6} & 0 & -1 \\
0 & \sqrt {6} & 0 &  - \sqrt {6} & 0 \\
1 & 0 &  - \sqrt {6} & 0 & 1 \\
0 & -1 & 0 & 1 & 0
\end{array}}\right],$$

$$Z_3=i  \left[ 
{\begin{array}{crrrc}
0 & 0 & 0 & 0 &  4\\
0 & 0 & 0 & 0 & 0 \\
0 & 0 & 0 & 0 & 0 \\
0 & 0 & 0 & 0 & 0 \\
 - 4\ & 0 & 0 & 0 & 0
\end{array}}
 \right]. $$

In a similar fashion to  the icosahedral case, 
we can consistently consider Nahm data of
the form $T_i(s)=x(s)\rho_i+y(s)Y_i$. The Nahm equations become
\begin{eqnarray}\frac{dx}{ds} &=& 2x^2-48y^2,\label{cog1}\\
\frac{dy}{ds} &=& -6xy-8y^2\label{cog2}\end{eqnarray}
and the spectral curve is
\be \eta^5+768\kappa^4\eta (\zeta^8+14\zeta^4+1)=0 \ee
where
\be \kappa^4=5y(x+3y)(x-2y)^2.\label{cog3}\ee
Equations (\ref{cog1}-\ref{cog2}) are identical to those for the
charge four cubic monopole \cite{HMM} and are solved by
\begin{eqnarray} x&=&\frac{2\kappa(5\wp^2(u)-3)}{5{\wpp}(u)},\\
y&=&\frac{2\kappa}{5{\wpp}(u)},\end{eqnarray}
where $u=2\kappa s$ 
and $\wp$ is the Weierstrass elliptic function 
satisfying
\be
\wpp^2=4(\wp^3-\wp).
\label{wp2}
\ee
 As in \cite{HMM}, the argument of $\kappa$ is
chosen to be $\pi/4$ and $u$ lies on the line from $0$ to
$\omega_2=\omega_1+\omega_3$, where $2\omega_1$ is the real period
of the elliptic function (\ref{wp2}) and
$2\omega_3$ is the imaginary period. By examining the eigenvalues of
the residue of $iT_3$ we see the boundary conditions 
at $s=0$ and $s=2$ are satisfied
provided,
\be\omega_2=4\kappa.\ee
This period may be explicitly calculated, with the result
that there exists an
octahedral monopole  with spectral curve
\be \eta^5+\frac{3\Gamma(\frac{1}{4})^8}
{16\pi^2}(\zeta^8+14\zeta^4+1)\eta=0.
\label{five}\ee

Note that the spectral curve (\ref{oct}) 
of the  cubic 4-monopole is
\be \eta^4+\Xi(\zeta^8+14\zeta^4+1)\eta=0 
\ee
for some constant $\Xi$ and the spectral curve (\ref{five}) 
of the octahedral 5-monopole is
\be
\eta\left[\eta^4+4\Xi(\zeta^8+14\zeta^4+1)\eta\right]=0
\ee
where $\Xi$ is the same constant.
The spectral curve of the octahedral 5-monopole  is therefore
given by a multiplication by $\eta$ of the cubic 4-monopole
spectral curve, up to the factor of $4$ in the constant.
Rather remarkably, this is exactly how the 
spectral curve of the axisymmetric 
3-monopole is obtained from that of the axisymmetric 2-monopole.
The two spectral curves in this case being \cite{HA}
\bea
\eta^2+\frac{\pi^2}{4}\zeta^2&=&0\label{twotorus}\\
\eta\left[\eta^2+\pi^2\zeta^2\right]&=&0.\label{threetorus}
\eea
The energy density of the 2-monopole described by (\ref{twotorus})
is axially symmetric, so that a surface of constant energy density
is toroidal. This is also true of the 3-monopole (\ref{threetorus})
and the only modification is that the torus is slightly larger in
size. This suggests that the octahedral 5-monopole may 
resemble a cube, since the cubic 4-monopole does so, with the
only modification being that the cube will be slightly larger.
The fact that equations (\ref{cog1}-\ref{cog2}) are identical
to those obtained in the cubic 4-monopole reduction of Nahm's
equations also supports this hypothesis.

Using our numerical scheme, we have calculated the energy
density of the octahedral 5-monopole. 
Fig. 2 shows a surface of constant energy density for this
monopole. It resembles an octahedron (not a cube) with
the energy density taking its maximum value on the six vertices of the
octahedron.
We found this result
quite surprising, given the comments above. However, it is good news
for our conjecture of Section 1, which claimed that this
monopole would look like an octahedron.

\begin{figure}[ht]
\begin{center}
\leavevmode
\epsfxsize=10cm \epsffile{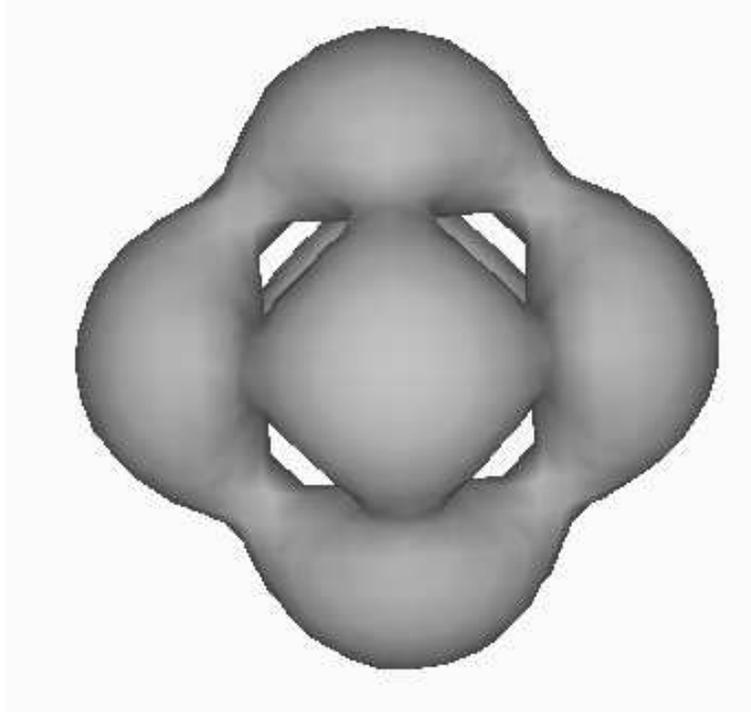}
\caption{Octahedral 5-monopole; surface of constant energy density
${\cal E}=0.07$.}
\end{center}
\end{figure}

\section{Rational Maps and Geodesic Scattering} 
\news
The $k$-monopole moduli space ${\cal M}_k$ is the space of gauge inequivalent
$k$-monopole
solutions to the Bogomolny equation (\ref{bogeq}). 
The motion of slow moving monopoles can be approximated by geodesic
motion in this moduli space \cite{M,S}. In this Section, we shall use
rational maps to present geodesics containing the octahedral
and dodecahedral monopoles.

A rational map of a $k$-monopole is a map from $\C$ to \CP$^1$ of the
form
\be R(z)=\frac{p(z)}{q(z)} \ee
where $q(z)$ is a
monic polynomial of degree $k$ and 
$p(z)$ is a polynomial of degree less than $k$, with no factors
in common with $q(z)$.
Donaldson has proved \cite{D} that
every rational map arises from
a unique $k$-monopole, so the space of such rational maps
 is diffeomorphic to ${\cal M}_k$. 

For our purposes, the most useful way to understand the relationship
between a monopole and its rational map is to follow the analysis of
Hurtubise \cite{Hu}. A line and an orthogonal plane in $\R^3$ are
chosen to give the decomposition
\be\R^3\cong\C\times\R.\ee
For convenience, we choose the line to be the $x_3$-axis and denote by $z$ the
complex coordinate on the $x_1x_2$-plane. Solutions to the linear
differential equation (\ref{de})
\be (D_A-i\Phi)v=0\ee
are considered along lines parallel to the $x_3$-axis. This equation has
two independent solutions. A basis $(v_0,v_1)$ for the solutions can be
chosen such that 
\begin{eqnarray} \lim_{x_3\rightarrow
    \infty}v_0(x_3)x_3^{\;-k/2}e^{x_3}&=&e_0,\\
   \lim_{x_3\rightarrow
    \infty}v_1(x_3)x_3^{\;k/2}e^{-x_3}&=&e_1\nonumber\end{eqnarray}
where $e_0$, $e_1$ are constant in some asymptotically flat
gauge. Thus $v_0$ is bounded and $v_1$ is unbounded as
$x_3\rightarrow\infty$. Similarly, there is a basis
$(v_0^{\prime},v_1^{\prime})$ such that $v_0^{\prime}$ is bounded and
$v_1^{\prime}$ is unbounded as $x_3\rightarrow-\infty$. We consider the
scattering along all lines and write
\begin{eqnarray} v_0^{\prime}&=&a(z)v_0+b(z)v_1,\\
  v_0&=&a^{\prime}(z)v_0^{\prime}+b(z)v_1^{\prime}.\end{eqnarray}
The rational map is given by
\be R(z)=\frac{a(z)}{b(z)}.\ee
Furthermore, since the spectral curve $P(\eta,\zeta)$ of a monopole corresponds
to the bounded solutions to (\ref{de}),
\be b(z)=P(z,0).\ee
Finally, it can be shown, \cite{AH} pp. 127-128, that the full
scattering data are given by
\be \left[\begin{array}{cc} a & b\\-b^\prime &
-a^{\prime}\end{array}\right]\left(\begin{array}{c}v_0\\v_1
\end{array}\right)\ =\left(\begin{array}{c}v_0^{\prime}\\v_1^{\prime}
\end{array}\right)\ee
where 
\be aa^{\prime}=1+b^\prime b.\ee

The advantage of rational maps is that monopoles
are easily described in this approach, since one simply
writes down any rational map. The disadvantage
is that the rational map tells us very little about the monopole. In
particular, since
the construction of the rational map requires the choice of a
direction in $\R^3$ it is not  possible to study the full symmetries
of a monopole from its rational map. However, the following isometries are
known \cite{HMM}. Let $\lambda\in U(1)$ and $\nu \in \C$ define a
rotation and translation respectively in the plane $\C$. Let $x\in\R$
define a translation perpendicular to the plane and let $\mu\in U(1)$
be a constant gauge transformation. Under the composition of these
transformation a rational map $R(z)$ transforms as
\be R(z)\rightarrow \mu^2e^{2x}\lambda^{-2k}R(\lambda^{-1}(z-\nu)).\label{trans1}\ee
Furthermore, under space inversion,
$x_3\rightarrow -x_3$, 
 $R(z)=p(z)/q(z)$ transforms as
\be \frac{p(z)}{q(z)}\rightarrow\frac{I(p)(z)}{q(z)}\label{trans2}\ee
where $I(p)(z)$ is the unique polynomial of degree less than $k$ such
that $(I(p)p)(z)=1$ mod $q(z)$. 

We note that this implies that the rational map of a charge $L$
axisymmetric monopole lying a distance $x$ above the plane is 
\be\frac{e^{2x+i\chi}}{z^L}\ee
and that the full scattering data for such a monopole are 
\be \left[ \begin{array}{cc}e^{2x+i\chi}&z^L\\0&-e^{-(2x+i\chi)}
\end{array} \right].\label{sdata}\ee

Using (\ref{trans1}) and (\ref{trans2}), it is easy to show that
(up to a choice of orientation) the most general rational map of
a strongly centred 5-monopole, which is invariant under both
inversion and $C_4$ rotation around the $x_3$ axis is
\be R_5(z)=\frac{\frac{2}{a}z^4+1}{z^5+az}\ee
with $a\in(0,\infty)$. It is a one
parameter family of based rational maps, corresponding to geodesic 
scattering of
5-monopoles. Since the octahedral monopole satisfies this symmetry, it
must lie on this geodesic.

Similarly, by imposing a $C_{10}$ symmetry on 7-monopoles,
generated by simultaneous inversion and rotation by
$\frac{\pi}{5}$, there is again a unique (up to orientation) one
parameter family of maps given by
\be R_7(z)=\frac{az^5+1}{z^7}\ee
with $a\in(-\infty,\infty)$. The dodecahedral monopole satisfies this
symmetry and must lie somewhere on the geodesic. 

We can understand these scattering processes by examining the rational maps
$R_5(z)$ and $R_7(z)$ for extreme values of
the parameter $a$. It is known \cite{HMM,Bi} that for a rational map $p(z)/q(z)$ with well
separated poles $\beta_1,\ldots,\beta_k$ the corresponding monopole is
approximately composed of unit charge monopoles located at the points
$(x_1,x_2,x_3)$, where $x_1+ix_2=\beta_i$ and 
$x_3=\frac{1}{2}\log{|p(\beta_i)|}$.
This approximation applies only when
the values of the numerator at the poles is small compared to the
distance between the poles.
 Thus, for large values of $a$, $R_5(z)$
corresponds to a monopole located at the origin and a monopole a
distance $\pm a$ along each of the diagonals $x_1=\pm x_2$.
This interpretation breaks down for $a\sim1$.
The poles of $R_7(z)$ are never well separated, so there is no
region in which this approximation can be applied to this rational
map.

In \cite{AH} pp. 25-26 it is argued that for monopoles strung out in well
separated clusters along, or nearly along, the $x_3$ axis the first
term
 in a large $z$
expansion of the rational map $R(z)$ will be $e^{2x+i\chi}/z^L$ 
where $L$ is the
charge of the topmost cluster and $x$ is its elevation above the
plane. We would like to extend this and argue that if the next
highest cluster has charge $M$ and is $y$ above the plane then the
first two terms in the large $z$ expansion of the rational map will be
given by
\be R(z)\sim\frac{e^{2x+i\chi}}{z^L}+\frac{e^{2y+i\phi}}{z^{2L+M}}+...
\label{paulsform}\ee

Assume the topmost cluster, $(A_1,\phi_1)$ is well separated from the other
monopoles. Let $v_0^{\prime\prime}$ be the solution bounded at
$x_3\rightarrow-\infty$. For $z$ large, we are considering scattering
along lines well removed from the spectral lines and so in the region
of $(A_1,\phi_1)$ the solution is dominated by the exponentially growing
one and is therefore close to $v_0^{\prime\prime}$. Thus the dominant term
in the rational map is the effect of scattering off $(A_1,\phi_1)$. 

We now consider the second highest monopole cluster $(A_2,\phi_2)$. Since it is
separated from the monopoles below it the incoming solution is close to
$v_0^{\prime\prime}$. If we call the bounded solution leaving the
$(A_2,\phi_2)$ region
$v_0^{\prime}$ and the unbounded one $v_1^{\prime}$ we have from
(\ref{sdata})
\be v_0^{\prime\prime}=e^{2y+i\phi}v_0^{\prime}+z^Mv_1^{\prime}.
\label{s0pp}\ee
Subsequent scattering off $(A_1,\phi_1)$ gives
\begin{eqnarray}v_0^{\prime}&=&-e^{-2x-i\chi}v_1\label{s0ps1p}\\
               v_1^{\prime}&=&e^{2x+i\chi}v_0+z^Lv_1\nonumber\end{eqnarray}
where $v_0$ and $v_1$ are respectively the unbounded and bounded
solutions as $x_3\rightarrow\infty$. Substituting (\ref{s0ps1p}) into
(\ref{s0pp}) we find that
\be v_0^{\prime\prime}=z^Me^{2x+i\chi}v_0+(z^{M+L}
-e^{-2(x-y)-i(\chi-\phi)})v_1\ee
and so the rational map is dominated by
\be R(z)\sim\frac{z^Me^{2x+i\chi}}{z^{M+L}-e^{-2(x-y)-i(\chi-\phi)}}\ee
and so since $x\gg y\gg 1$
\be R(z)-\frac{e^{2x+i\chi}}{z^L}\sim\frac{e^{2y+i\phi}}{z^{2L+M}}\ee
as required. Obviously this type of argument could be extended to
further monopoles along the line, but we do not need to do so here.

We can now see that $R_5(z)$ describes four monopoles approaching
a monopole at the origin along the negative and positive directions of
the $x_1$ and $x_2$ axis. At some point, the monopoles coalesce to form
the octahedral 5-monopole.
 As $a\rightarrow0$, we see from
(\ref{paulsform}) that one monopole travels up the $x_3$-axis and three
remain in a cluster at the origin. By inversion the fifth monopole
travels down the $x_3$-axis. In the $a=0$ limit, there
are spherical unit charge monopoles at $(0,0,\pm\infty)$ and a
 toroidal 3-monopole centred on the origin. 

Similarly the rational map $R_7(z)$ corresponds to two 2-monopole
clusters approaching a toroidal 3-monopole along the positive and negative
$x_3$-axis. At some negative value of $a$, say $a=-d$, they coalesce
 to form a dodecahedron oriented so that two
faces are parallel to the $x_1x_2$-plane.  Then at $a=0$ they form a
toroidal 7-monopole. At $a=d$ they form another dodecahedron, rotated
$\pi/5$ relative to the previous one. Finally, for large values of $a$ the
rational map corresponds to a toroidal 3-monopole at the origin and
two 2-monopole clusters receding along the positive and negative $x_3$-axis.

 Recently, we have been
investigating a whole family of scattering geodesics similar
to the one above. 
One of the interesting features of these scattering  processes is
the complicated motion of the zeros of the Higgs field.
A detailed investigation will be presented elsewhere
\cite{HSc}.

\section{Conclusion}
\news
By explicit construction of the spectral curves, we
have proved the existence of a charge seven monopole with
icosahedral symmetry and a charge five monopole with octahedral
symmetry. Numerical computation of the monopole energy density
reveals that the former looks like a dodecahedron and the latter
an octahedron. The energy density is maximal on the vertices
of these two regular solids.

Using Donaldson's rational map formulation we have presented a totally 
geodesic one-dimensional submanifold of the 
monopole moduli space which contains the dodecahedral
7-monopole and one which contains the octahedral 5-monopole.
In the moduli space approximation of soliton dynamics, these 
submanifolds describe a new type of novel multi-monopole 
scattering which requires further investigation.\\

\noindent{\bf Acknowledgements}

Many thanks to Nigel Hitchin and Nick Manton for useful 
discussions. CJH thanks the EPSRC for a research studentship and the
British Council for a FCO award. PMS thanks the EPSRC
for a research fellowship.\\

\end{document}